\documentclass[conference]{IEEEtran}
\usepackage{amssymb,amsfonts,amsmath,amsthm,amscd,dsfont,mathrsfs}
\usepackage{graphicx}
\DeclareMathAlphabet{\mathpzc}{OT1}{pzc}{m}{it}

\newtheorem{propo}{Proposition}[section]
\newtheorem{lemma}[propo]{Lemma}
\newtheorem{definition}[propo]{Definition}
\newtheorem{hypothesis}[propo]{Hypothesis}

\newcommand{\bitem}{\begin{itemize}}
\newcommand{\eitem}{\end{itemize}}

\newcommand{\<}{\langle}
\newcommand{\Alg}{{\cal A}}
\newcommand{\plim}{\mbox{\rm p.lim}}
\newcommand{\LSLim}{\mbox{\rm ls.lim}}
\newcommand{\goto}{\rightarrow}
\renewcommand{\>}{\rangle}

\newcommand{\reals}{{\mathds R}}

\newcommand{\naturals}{{\mathds N}}
\newcommand{\cF}{{\cal F}}
\def\prob{{\mathds P}}

\def\E{\mathds E}
\def\1{\mathds 1}
\def\de{{\rm d}}

\def\var{{v}}

\def\hsigma{\widehat{\sigma}}

\def\BPDN{{\sf BPDN}}

\def\AMP{{\sf AMP}}
\def\AMPM{{\sf AMP.M}}
\def\AMPA{{\sf AMP.A}}
\def\AMPT{{\sf AMP.T}}
\def\AMP0{{\sf AMP.0}}
\def\HFP{{\rm HFP}}
\def\SC{{\rm SC}}
\def\EqDR{{\rm EqDR}}

\begin{document}

\title{Message Passing Algorithms for Compressed Sensing:
II. Analysis and Validation}

\author{\IEEEauthorblockN{David L. Donoho}
\IEEEauthorblockA{Department of Statistics\\
Stanford University
}
\and
\IEEEauthorblockN{Arian Maleki}
\IEEEauthorblockA{Department of Electrical Engineering\\
Stanford University}
\and
\IEEEauthorblockN{Andrea Montanari}
\IEEEauthorblockA{Department of Electrical Engineering\\
and Department of Statistics\\
Stanford University
}}

\maketitle

\begin{abstract}
In a recent paper, the authors proposed a new class of low-complexity
iterative thresholding algorithms for reconstructing sparse signals
from a small set of linear measurements \cite{DMM}.
The new algorithms are broadly referred to as
AMP, for \emph{approximate message passing}.
This is the second of
two conference papers describing the derivation of
these algorithms, connection with related literature,
extensions of original framework, and new empirical evidence.

This paper describes the state evolution formalism for analyzing
these algorithms, and some of the conclusions that can be
drawn from this formalism. We carried out extensive
numerical simulations to confirm these predictions.
We present here a few representative results.
\end{abstract}

\IEEEpeerreviewmaketitle

\section{General AMP and State Evolution}

We consider the model
\begin{eqnarray}
y = A\, s_o + w_o\, ,\, \;\;\; s_o\in\reals^N,\; y, w_o \in\reals^n\,,
\end{eqnarray}
with $s_o$ a vector that is `compressible' and $w_o$
a noise vector.  We will assume that the entries of
$w_o$ are centered independent gaussian random variables with
variance $\var$.

The general AMP (approximate message passing) algorithm reads
\begin{eqnarray}
x^{t+1} & = & \eta_t(x^t+A^*z^t)\, ,\\
z^t & = & y -Ax^t + \frac{1}{\delta}z^{t-1}
\<\eta'_{t-1}(x^{t-1}+A^*z^{t-1})\>
\, ,
\end{eqnarray}
with initial condition $x_0=0$.
Here,
for a vector $u=(u_1,\dots,u_N)$ we write
$\<u\> \equiv \sum_{i=1}^N u_i/N$, and $\eta'(\,\cdot\,;\,\cdot\,)$
indicates the derivative of $\eta$ with respect to its first argument.
Further $\delta \equiv n/N$ and $\{\eta_t(\,\cdot\,)\}_{t\ge 0}$ is a sequence
of scalar non-linearities (see Section III), a typical example being
soft thresholding, which contracts its argument towards zero.

\subsection{Structure of the Algorithm}

This algorithm is interesting for its low complexity: its implementation
is dominated at each step by the cost of applying $A$ and $A^*$ to
appropriate vectors.  In some important settings, matrices $A$ of interest
can be applied to a vector implicitly by a pipeline of
operators requiring $N \log(N)$
flops; an example would be $A$ whose rows are randomly chosen from
among the rows of a Fourier matrix; then $Ax$ can be computed by FFT
and subsampling.

Even more, the algorithm is interesting for the message passing term
$ \frac{1}{\delta}z^{t-1} \<\eta'_{t-1}(x^{t-1}+A^*z^{t-1})\>$.
Similar algorithms without this term are common in the literature
of so-called iterative thresholding algorithms.  As discussed
in the companion paper, the message passing term
approximates the combined effect on the reconstruction of
the passing of $nN$ messages in the the full
message passing algorithm.

The message passing term completely
changes the statistical properties of the reconstruction, and
it also makes the algorithm amenable to analysis by a
technique we call State Evolution.  Such analysis
shows that the algorithm converges rapidly, much more
rapidly than any known result for the IST algorithm.
Furthermore, it allows us to make a variety of theoretical
predictions about performance characteristics of the algorithm
which are much stronger than any predictions available
for competing methods.

\subsection{State Evolution}

In the following we will assume that the columns
of $A$ are normalized to unit Euclidean length.
We define the {\it effective variance}
\begin{eqnarray}
\sigma(x_t)^2\equiv \var+\frac{1}{N\delta}\, ||x_t-s_0||_2^2\, .
\end{eqnarray}
The effective variance combines the observational variance $v$
with an additional term $\frac{1}{N\delta}\, ||x_t-s_0||_2^2$
that we call the interference term. Notice that $v$
is merely the squared reconstruction error of the naive `matched filter'
for the case where $s_0$ contains all zeros and a single nonzero
in a given position $i$ and the matched filter is just
the $i$-th column of $A$.

The  interference term measures the additional
error in estimating a single component of $s_{o,i}$ that is caused
by the many small errors in {\it other components} $j \neq i$.
The formula states that the effective variance at iteration $t$
is caused by the observational noise (invariant across
iteration) and the current errors
at iteration $t$ (changing from iteration to iteration).
The interference concept is well known in digital communications,
where phrases like {\it mutual access interference} are used
for what is algebraically the same phenomenon.

We will let $\hsigma_t$ denote any estimate of $\sigma_t$,
and we will assume that $\hsigma_t\approx \sigma_t$;
see \cite{DMM} for more careful discussion.
Suppose that the nonlinearity takes the form
$\eta_t(\, \cdot\, ) = \eta(\, \cdot\,
;  \theta_t)$
where $\theta$ is a tuning parameter, possibly depending on $\hsigma_t$;
see below for more. Let $\cF$ denote the collection of
CDFs on $\reals$ and $F$ be the CDF of $s_0(i)$.
Define the MSE map $\Psi: \reals^+ \times  \reals^3 \times \cF \mapsto \reals^+$ by
\[
\Psi( \sigma ; v , \delta,  \theta_t, F)  = \var+\frac{1}{\delta}\,
\E\Big\{\big[\eta_t\big(X+\sigma\, Z\big) -X\big]^2\Big\}\, \label{eq:GeneralSE}
\]
where $X$ has distribution $F$ and $Z \sim \mathcal{N}(0,1)$ is independent of $X$. We suppose that a rule
$\Theta(\sigma; v, \delta, \theta,F)$ for the update
of $\theta_t$ is also known.
\begin{definition}
{\sl The {\em state} is a 5-tuple $S = (\sigma; v,\delta, \theta,F) $;
{\em state evolution} is  the evolution of the state by the rule
\begin{eqnarray*}
(\sigma_t^2; v,\delta, \theta_t,F) &\mapsto& (\Psi(\sigma_t^2); v,\delta,  \theta_{t+1},F)\\
 t  &\mapsto&  t+1
\end{eqnarray*}
As the parameters  $(v,\delta,F)$ remain
fixed during evolution, we usually omit mention of them and think of
state evolution simply as the  iterated application of $\Psi$ and $\Theta$:
\begin{eqnarray*}
\sigma_t^2&\mapsto& \sigma_{t+1}^2 \equiv \Psi(\sigma_t^2)\\
\theta_t&\mapsto& \theta_{t+1} \equiv \Theta(S_t)\\
 t  &\mapsto&  t+1
\end{eqnarray*}
The initial state is taken to have $\sigma_0^2=\var+ ||s_0||_2^2/N\delta$.
}
\end{definition}
\newcommand{\cE}{{\cal E}}
As described, State Evolution is a purely analytical
construct, involving sequential application of
rules $\Psi$ and $\Theta$.  The crucial point is to know
whether this converges to a fixed point, and
to exploit the properties of the fixed point.
We expect that such properties are reflected
in the properties of the algorithm.  To make this
precise, we need further notation.
\begin{definition} {\bf State-Conditional Expectation}.
Given a function $\zeta: \reals^4 \mapsto \reals$,
its \emph{expectation in state $S_t$} is
\[
      \cE(\zeta | S_t) = \E\,\big\{ \zeta(U,V,W,\eta(U+V+W)) \big\}\, ,
\]
where $U \sim F$, $V \sim N(0,v)$ and $W \sim N(0,\sigma_t^2 - v)$.
\end{definition}
Different choices of $\zeta$ allow to monitor the
evolution of different metrics under the AMP algorithm.
For instance, $\zeta = (u-x)^2$
corresponds to the mean square error (MSE).
The False Alarm Rate is tracked by $\zeta = 1_{\{ \eta( v+w) \neq 0 \}}$
and the Detection Rate by $\zeta = 1_{\{ \eta( u+ v+w) \neq 0 \}}$.

\begin{definition} {\bf Large-System Limits}.
Let $\zeta: \reals^4 \mapsto \reals$ be a  function
of  real 4-tuples $(s, u, w, x)$.
Suppose we run the iterative algorithm $\Alg$ for a sequence of problem
sizes $(n,N)$  at a the value $(v,\delta,F)$ of underlying implicit
parameters, getting outputs $x_t$, $t = 1, 2 , 3 , \dots $
The large-system limit $\LSLim(\zeta,t,\Alg)$
of $\zeta$ at iteration $t$ is
\[
   \LSLim(\zeta,t,\Alg) = \plim_{N \goto \infty} \< \zeta(s_{o,i},u_{t,i},w_{o,i},x_{t,i}) \>_N\, ,
\]
where $\<\,\cdot\,\>_N$ denotes the uniform average over $i\in\{1,\dots,N\}\equiv[N]$, and $\plim$ denotes limit in probability.
\end{definition}

\begin{hypothesis}
{\bf Correctness of State Evolution for AMP}.
Run an AMP algorithm  for $t$ iterations with 
implicit state variables $v,\delta, F$.
Run state evolution, obtaining the state $S_t$ at time $t$.
Then  for any bounded continuous function 
$\zeta: \reals^4 \mapsto \reals$ of the real 4-tuples $(s, u, w, x)$,
and any number of iterations $t$,
\begin{enumerate}
\item The large-system limit $\LSLim(\zeta,t,\Alg)$
exists for the observable $\zeta$ at iteration $t$.
\item This limit coincides with the 
expectation $\cE(\zeta|S_t)$ computed at state $S_t$.
\end{enumerate}
\end{hypothesis}

\vspace{0.1cm}

State evolution, where correct, allows us to predict the
performance of AMP algorithms and tune them for optimal performance.
In particular, SE can help us to choose the 
non-linearities $\{\eta_t\}$ and their tuning. 
The objective of the rest of this paper is twofold:
$(1)$ Provide evidence for state evolution;
$(2)$ Describe some guidelines towards the choice of
the non-linearities $\{\eta_t\}$.

\section{AMP-Based algorithms}

Already in \cite{DMM} we showed that a
variety of algorithms can be generated by
varying the choice of $\eta$.  We begin
with algorithms based on soft thresholding.
Here $\eta_t(x) =\eta(x;\theta_t)$ is given
by the soft threshold function
\newcommand{\eps}{\epsilon}
\begin{eqnarray}
\eta(x;\theta) = \left\{\begin{array}{ll}
x-\theta& \mbox{ if $\theta<x$,}\\
0 & \mbox{ if $-\theta\le x\le\theta$,}\\
x+\theta  & \mbox{ if $x<-\theta$.}
\end{array}\right.
\end{eqnarray}
This function shrinks its argument towards the origin.
Several interesting AMP-Based algorithms are obtained
by varying the choice of the sequence
$\{\theta_t\}_{t\in\naturals}$.

\subsection{$\AMPM(\delta)$}

The paper \cite{DMM} considered the noiseless
case $v=0$ where the components of $s_o$ are iid 
with common distribution $F$
that places all but perhaps a fraction $\eps  = \rho(\delta) \cdot \delta$,
$\rho \in (0,1)$ of its mass
at zero. That paper proposed the choice
\begin{eqnarray}
\theta_t = \tau(\delta)\hsigma_t\, .
\end{eqnarray}
where an explicit formula for $\tau(\delta)$ is derived
in the online supplement \cite{DoMaMo09Supplement}. As explained
in that supplement, this rule has a minimax interpretation,
namely, to give the smallest MSE guaranteed
across all distributions $F$ with mass 
at zero larger than or equal to $1-\eps$.

%
%
\subsection{$\AMPT(\tau)$}

Instead of taking a worst case viewpoint, we can
think of specifically tuning for the case at hand.
Consider general rules of the form:
\begin{eqnarray}
\theta_t = \tau\, \hsigma_t\, .
\end{eqnarray}
Such rules  have a very convenient property
for state evolution; namely, if we suppose that $\hsigma_t \equiv \sigma_t$,
we can redefine the
state as $(\sigma_t^2 ; v , \delta, \tau, F)$, with $(v , \delta, \tau, F)$
invariant during the iteration,
and then the evolution is effectively one-dimensional:
$\sigma_t^2 \mapsto \sigma_{t+1}^2 \equiv \Psi(\sigma_t^2)$.
The dynamics are then very easy to study, just by looking for fixed
points of a scalar function $\Psi$.
(This advantage is also shared by $\AMPM(\delta)$, of course).

While the assumption $\hsigma_t \equiv \sigma_t$ does not hold, 
strictly speaking, at any finite size, it will hold 
asymptotically in the large system limit for many good estimators of the 
effective variance.

It turns out that, depending on $F$ and $\delta$,
different values of $\tau$ lead to very different performance characteristics.
It is natural to ask for the fixed value $\tau  = \tau^*(v,\delta,F)$
which, under state evolution gives the smallest
equilibrium MSE.  We have developed software
to compute such optimal tuning; results are
discussed in  \cite{LassoOpChar}.
%
%
\subsection{$\AMPA(\lambda)$}

In much current work on compressed sensing,
it is desired to solve the $\ell_1$-penalized
least squares problem
\begin{equation} \label{BPDN}
  \mbox{minimize }\;\; \frac{1}{2}\| y - A x \|_2^2 + \lambda \| x\|_1 .
\end{equation}
In different fields this has been called Basis Pursuit denoising \cite{ChDoSa98}
or Lasso \cite{LASSO}.
Large scale use of general convex solvers is impractical
when $A$ is of the type interesting from compressed sensing,
but AMP-style iterations {\it are} practical.
And, surprisingly an AMP-based algorithm can effectively
compute the solution by letting the threshold `float' to find the
right level for solution of the above problem. The threshold recursion is:
\begin{eqnarray}
\theta_{t+1} & = & \lambda + \frac{\theta_t}{\delta}
\<\eta'(x^{t}+A^*z^{t};\theta_{t})\>\, .
\end{eqnarray}
%
%
%
\subsection{$\AMP0$}

It can also be of interest to solve the $\ell_1$-minimization
problem
\begin{equation} \label{BP}
    \min_x  \| x\|_1  \mbox{ subject to } y = Ax .
\end{equation}
This has been called Basis Pursuit \cite{ChDoSa98}
in the signal processing literature. While
formally it can be solved by linear programming,
standard linear program codes are far too slow for
many of the applications  interesting to us.

This is formally the $\lambda=0$ case of $\AMPA(\lambda)$.
In fact it can be advantageous to allow $\lambda$
to decay with the iteration number
\begin{eqnarray}
\theta_{t+1} & = & \lambda_t + \frac{\theta_t}{\delta}
\<\eta'(x^{t}+A^*z^{t};\theta_{t})\>\, .
\end{eqnarray}
Here, we let $\lambda_t\downarrow 0$ as $t\to\infty$.
%
%

%
%
%


%
%

\subsection{Other Nonlinearities}

The discussion above has focused entirely on
soft thresholding, but both the AMP algorithm
and SE formalism make perfect sense with many other nonlinearities.
Some case of specific interest include
\bitem
  \item  The Bayesian conditional mean: $\eta(x) = 
\E \{ s_0 | s_0 + U + V = x\}$,
  where $U$ and $V$ are just as in Definition I.2.
This is indeed discussed in the companion paper \cite{DoMaMo09itw},
Section V.
  \item  Scalar nonlinearities associated to various nonconvex 
optimization problems, such as minimizing $\ell_p$ pseudonorms for
$p<1$.
 \eitem

\section{Consequences of State Evolution}

\subsection{Exponential  Convergence of the Algorithm}

When State Evolution is correct for an AMP-type algorithm,
we can be sure that the algorithm converges rapidly to its
limiting value -- exponentially fast.
The basic point was shown in \cite{DMM}.
Suppose we are considering either $\AMPM(\delta)$
or $\AMPT(\tau)$. In either case, as explained above,
the state evolution is effectively one-dimensional. Then
the following is relevant.

\begin{definition}
{\bf Stable Fixed Point}.
The Highest Fixed Point of the continuous function $\Psi$ is
\[
    \HFP(\Psi) = \sup\{ m : \Psi(m) \geq m \}.
\]
The stability coefficient of the continuously differentiable function
$\Psi$ is
\[
    \SC(\Psi) = \left.\frac{\de\phantom{m}}{\de m} \Psi(m) \right|_{m = \HFP(\Psi)} \, .
\]
We say that $\HFP(\Psi)$ is a stable fixed point if $0 \leq \SC(\Psi) < 1$.
\end{definition}

Let $\mu_2(F) = \int x^2 dF$ denote the second-moment functional
of the CDF $F$.
\begin{lemma}
Let $\Psi(\,\cdot\,) = \Psi(\,\cdot\,; v,\delta, F)$.
Suppose that $\mu_2(F) > \HFP(\Psi)$.
The sequence of iterates $\sigma^2_t$ defined by
starting from $\sigma^2_0 = \mu_2(F)$ and $\sigma^2_{t+1}= \Psi(\sigma^2_{t})$
converges:
\[
           \sigma^2_t \goto \HFP(\Psi) , \qquad t \goto \infty.
\]
Suppose that the stability coefficient $0 < \SC(\Psi) < 1$. Then
\[
          (\sigma^2_t - \HFP(\Psi)) \leq \SC(\Psi)^t  \cdot (\mu_2(F) - \HFP(\Psi)) .
\]
\end{lemma}

In short, when $F$ and $v$ are such that
the highest fixed point is stable,
 state evolution converges exponentially fast
to that fixed point.

Other iterative thresholding algorithms have theoretical
guarantees which are far weaker. For example,
FISTA \cite{Fista} has a theoretical guarantee of $O(1/t^2)$,
while SE evolution implies $O(\exp(-ct))$.

\subsection{Phase Transitions For $\ell_1$ minimization}

Consider the special setting where the noise is absent $w_o = 0$
and the object $s_o$
obeys a strict sparsity condition;  namely the distribution
$F$ places a fraction $\geq 1-\eps$ of its mass at the
origin; and thus, if $s_o$ is iid $F$, approximately $N \cdot (1-\eps)$
of its entries are exactly zero.

A phase transition occurs in this setting when using $\ell_1$
minimization for reconstruction.  Namely,  if we write $\eps = \rho \cdot \delta$
then there is a critical value $\rho(\delta)$ such that, for $\eps < \rho(\delta) \cdot \delta$,
$\ell_1$ minimization correctly recovers $s_o$, while for $\eps >  \rho(\delta) \cdot \delta$,
$\ell_1$ minimization fails to correctly recover $s_o$,
with probability approaching one in the large size limit.
State Evolution predicts this phenomenon, because, for $\eps <  \rho_{\rm SE}(\delta) \cdot \delta$,
the highest fixed point is at $\sigma_t^2 = 0$, while above this value,
the highest fixed point is at $\sigma_t^2 > 0$.
Previously, the exact critical value $\rho(\delta)$ at which this transition occurs
was computed by combinatorial geometry, with a rigorous proof; however, it was shown in
\cite{DMM} that the algorithm $\AMPM(\delta)$ has $\rho(\delta) = 
\rho_{\rm SE}(\delta)$,
validating the correctness of SE.

\subsection{Operating Characteristics of $\ell_1$ penalized Least-squares.}

State evolution predicts the following relationships between $\AMPT(\tau)$ and
$\BPDN(\lambda)$.
$\AMPT(\tau)$ has, according to SE, for its large-$t$ limit
an {\it equilibrium state} characterized by its equilibrium noise plus interference level
$\sigma_\infty(\tau)$. In that state $\AMPT(\tau)$ uses an
equilibrium threshold $\theta_\infty(\tau)$.  Associated to this equilibrium
NPI and Threshold, there is an equilibrium detection rate
\[
      \EqDR(\tau) = \prob \{  \eta(U + V + W; \theta_\infty)  \neq 0 \}
\]
where $U \sim F$, $V$ is $N(0,v)$ and $W$ is $N(0,\sigma^2_\infty - v)$,
with $U$,$V$,$W$ independent.
Namely, for all sufficiently large $\tau$ (i.e $\tau > \tau_0(\delta,F,v)$) we have
\[
    \lambda = (1  - \EqDR(\tau)/\delta) \cdot \theta_\infty(\tau);
\]
this creates a one-one relationship $\lambda \leftrightarrow \tau(\lambda; v, \delta, F)$
calibrating the two families of procedures.
SE predicts that observables of the $\ell_1$-penalized least squares
estimator with penalty $\lambda$ will agree with the calculations
of expectations for $\AMPT(\tau(\lambda;v,\delta, F))$
made by state evolution.

\section{Empirical Validation}

The above-mentioned consequences of State Evolution can be tested as follows.
In each case, we can use SE to make a fixed prediction in advance of an experiment and then
we can run a simulation experiment to test the accuracy of the prediction.

\subsection{SE Predictions of Dynamics of Observables}

Exponential convergence of AMP-based algorithms is equivalent to
saying that a certain observable -- Mean-squared error of reconstruction --
decays exponentially in $t$.   This is but one observable of the algorithm's
output; and we have tested not only the SE predictions of MSE but
also the SE predictions of many other quantities.

\begin{figure}
\includegraphics[height=2.5in]{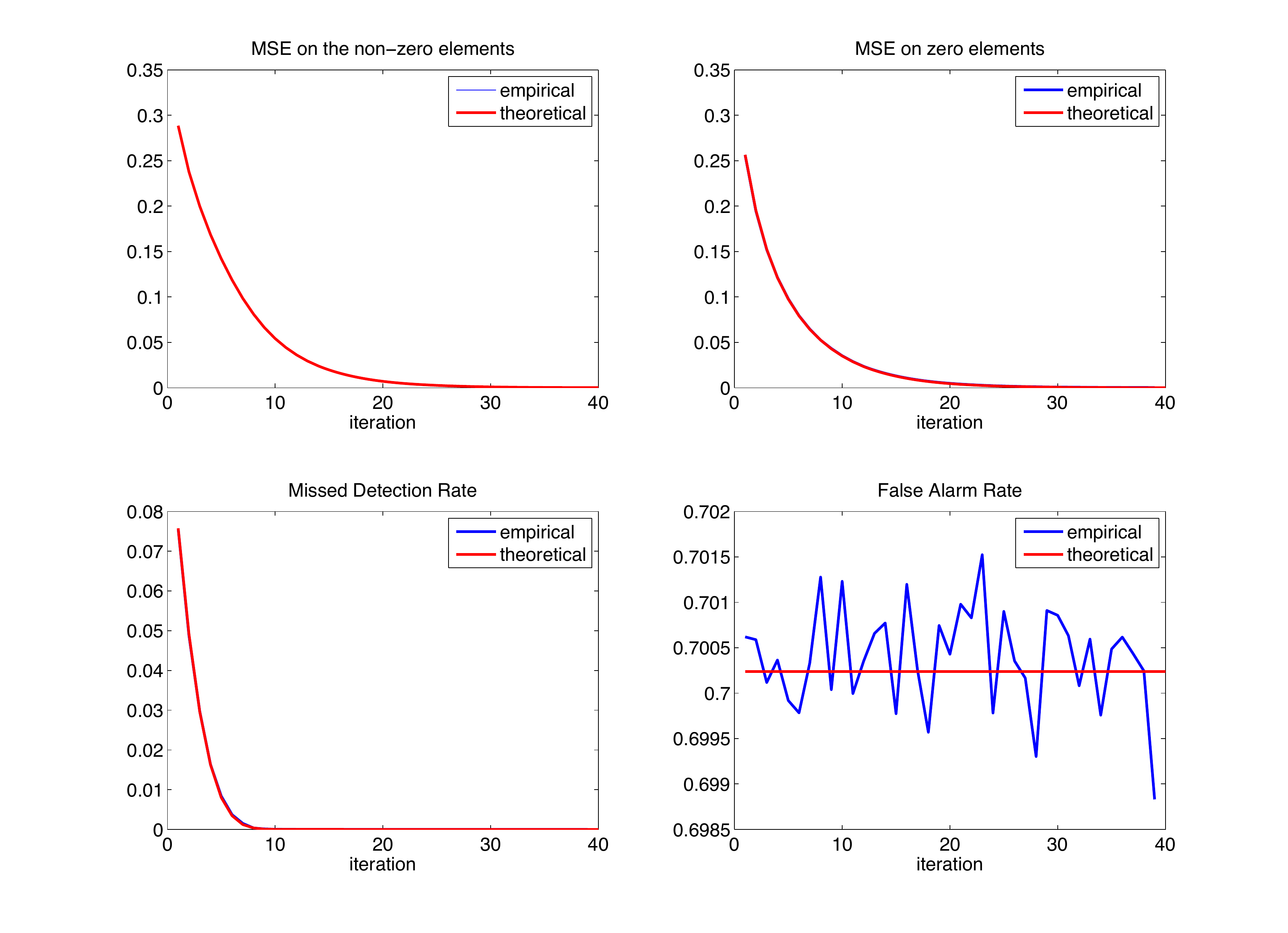}
\caption{Observables versus iteration, and predictions by state evolution.
Panels (a)-(d):  MSENZ, MSE, MDR, FAR.
Curve in red: theoretical prediction. Curve in blue:
mean observable. For this experiment,
$N=5000$,  $\delta= n/N = .3$. $F = 0.955 \delta_{0} + 0.045 \delta_{1}$}
\label{Fig-Observables}
\end{figure}

In Figure \ref{Fig-Observables} we present results from an experiment
with signal length $N=5000$,
noise level $v=0$, indeterminacy $\delta = n/N = 0.30$
and sparsity level $\eps = 0.045$. The distribution $F$ places $95.5\%$
of its mass at zero and $4.5\%$ of its mass at $1$.
the fit between predictions and observations is
extremely good -- so much so that it is hard to tell the two curves apart.
 For more details, see \cite{DoMaMo09Supplement}.

\subsection{Phase Transition Calculations}

Empirical observations of Phase transitions  of $\ell_1$ minimization
and other algorithms have been made in \cite{MaDo09SP,DoTa09},
and we follow a similar procedure. Specifically,
to observe a phase transition in the performance
of a sparsity-seeking algorithm, we perform 200
reconstructions on randomly-generated problem
instances with the same underlying situation ($v = 0$,
$\delta$, $F$) and we record the fraction of successful reconstructions
in that situation.  We do this for each member of a large set
of situations by varying the undersampling ratio $\delta$ and
varying sparsity of $F$.  More specifically, we define a
$(\delta,\rho)$ phase diagram $[0,1]$ and consider
a grid of sites in this domain with $\delta = .05, .10 , \dots$
and $\rho = .03, .06, \dots, .99$.  For each $\delta,\rho$ pair
in this grid, we generate random problem instances having
a $k$-sparse solution $s_0$, i.e. a vector
having $k$ ones and $n-k$ zeros;
here k = $\rho \cdot \delta \cdot N$.

Defining success as exact recovery of $s_0$ to within a small fixed
error tolerance, we define the empirical phase transition as
occurring at the $\rho$ value where the success fraction drops below 50\%.
For more details, see \cite{DoMaMo09Supplement}.

Figure 2 depicts the theoretical phase transition predicted by State Evolution
as well as the empirical phase transition of $\AMPM(\delta)$
and a traditional iterative soft thresholding algorithm. In this
figure $N = 1000$, and $\AMPM(\delta)$ was run for $T =1,000$
iterations.   One can see that empirical phase transition of
$\AMPM(\delta)$ matches closely the state evolution prediction.
One can also see that the empirical phase transition of iterative soft thresholding,
without the message passing term, is substantially worse than that
for the AMP-based method with the message passing term.

\begin{figure}
\includegraphics[height=2.5in]{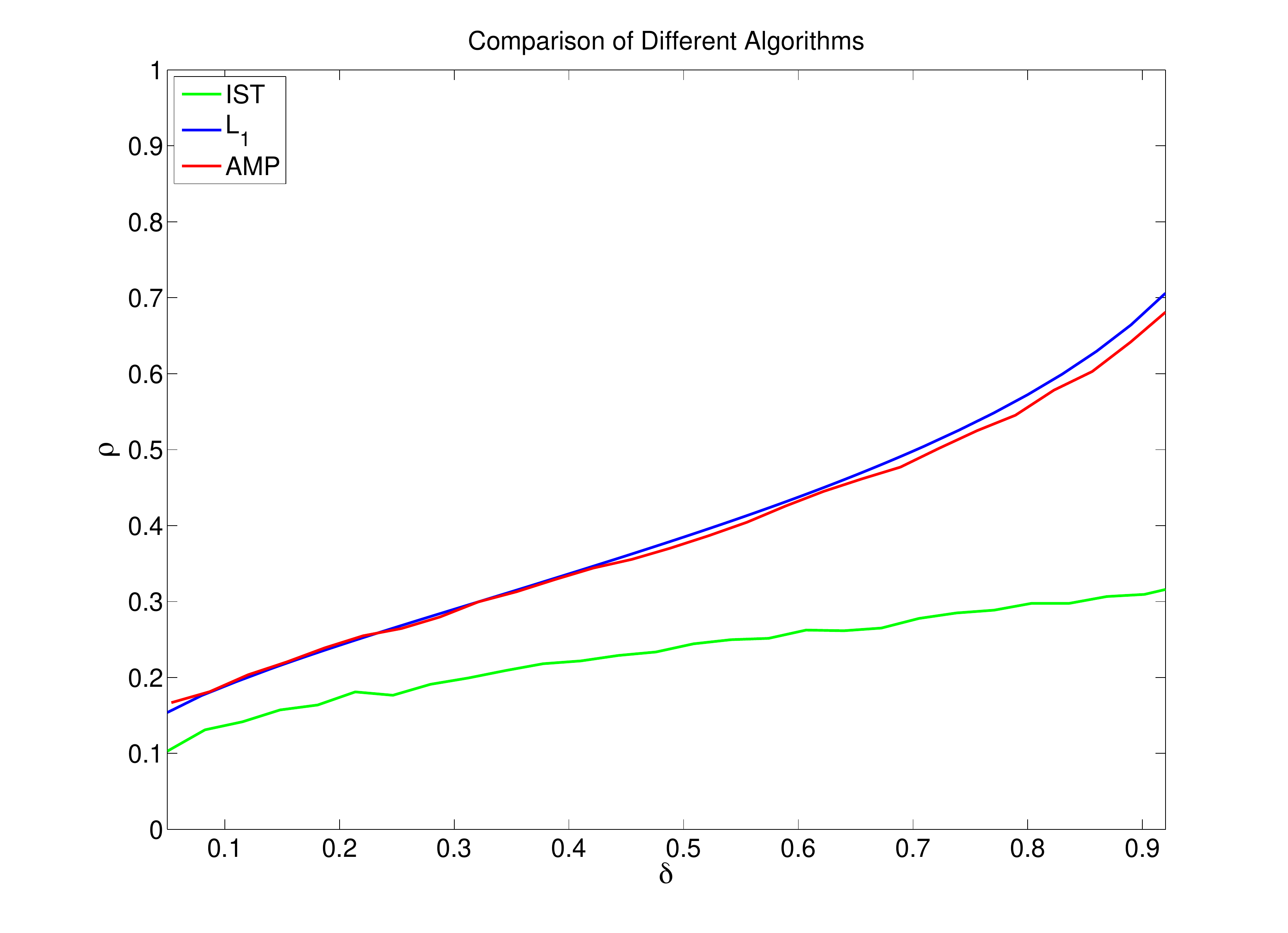}
\caption{Phase transitions of reconstruction algorithms.
Blue Curve:  Phase Transition predicted by SE; Red Curve empirical phase transition
for $AMPM(\delta)$  as observed in simulation; Green Curve,
empirical phase transition for Iterative Soft Thresholding
as observed in simulation.}
\label{fig2}
\end{figure}

\subsection{Operating Characteristics of $\ell_1$ penalized Least-squares}

The phase transition study gives an example of SE's accuracy
in predicting AMP-based algorithms in a {\it strictly} sparse setting,
i.e. where only a small fraction of entries in $s_0$ are nonzero.
For a somewhat different example, we consider the generalized
Gaussian family, i.e. distribution functions $F_\alpha$ with densities
\[
      f_\alpha(x) = \exp( -  |x|^\alpha )/Z_\alpha . 
\]
In the case $\alpha =1$ there is a very natural connection with $\ell_1$-minimization
algorithms, which then become MAP estimation schemes.  In the case
$\alpha=1$, an iid realization from $f_\alpha$, properly rescaled to
unit $\ell_1$ norm, will be uniformly distributed on the surface of the $\ell_1$ ball,
and in that sense this distribution samples all of the $\ell_1$ ball, unlike
the highly sparse distributions used in the phase transition study, which sample 
only the low-dimensional faces. When $\alpha < 1$, the sequence is
in a sense more sparse than when $\alpha = 1$.  The case $\alpha = .7$
has been found useful in modelling wavelet coefficients of natural images.

We considered exponents $\alpha \in \{ 0.35, 0.50, 0.65, 0.75, 1.0 \}$.
At each such case we considered incompleteness ratios 
 $\delta \in \{ 0.1, 0.2, 0.3, 0.4, 0.5 \}$.  The set of resulting $(\alpha,\delta)$ pairs
 gives a collection of 25 experimental conditions.  At each such experimental
 condition, we considered  5 or so different values of $\lambda$
 for which SE-predicted MSE's were available.  In total,
 simulations were run for 147 different combinations of $\alpha$, $\delta$
 and $\lambda$.  At each such combination, we randomly generated 200 problem
 instances using the problem specification, and then computed more than 50
 observables of the solution.  In this subsection, we used $N=500$.
 
 To solve an instance of problem (\ref{BPDN}) we had numerous options.
 Rather than a general convex optimizer, we opted to use
 the LARS/LASSO algorithm.  
  
 Figure 3 shows a scatterplot comparing MSE values for the LARS/LASSO
 solution of (\ref{BPDN}) with predictions by State Evolution, as decribed in section III.C.
 Each data point corresponds to one experimental combination of $\alpha$,
 $\delta$, $\lambda$, and the datapoint presents the median MSE across
 200 simulations under that combination of circumstances. 
 Even though the observed MSE's vary by more than an order of magnitude,
 it will be seen that the SE predictions track them accurately.  It should be
recalled that the problem size here is only $N=500$, and that only $200$
replications were made at each experimental situation.  In contrast,
the SE prediction is designed to match large-system limit.
In a longer paper, we will consider a much wider
range of observables and demonstrate that, at larger problem sizes $N$,
we get successively better fits between observables and their SE
predictions.  

\begin{figure}
\includegraphics[height=2.5in]{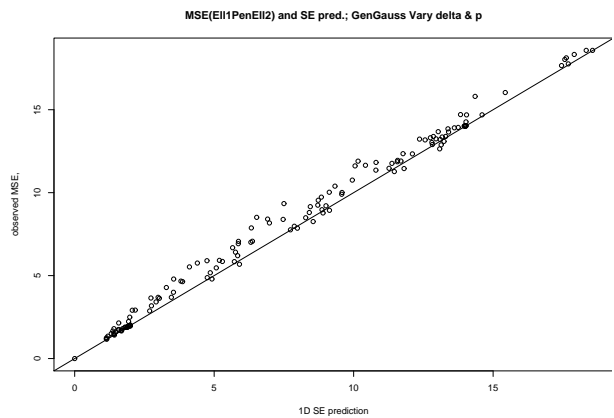}
\caption{Mean-squared Error for $\ell_1$-penalized Least-squares estimate
versus predicted error according to State Evolution.}
\end{figure}

\section*{Acknowledgment}

The authors would like to thank NSF for support in grants
DMS-05-05303 and DMS-09-06812, and
 CCF-0743978 (CAREER) and DMS-0806211 (AM).

\end{document}